\documentclass[a4paper,fleqn,usenatbib]{mnras}

\usepackage[T1]{fontenc}
\usepackage{ae,aecompl}

\usepackage{graphicx}	% Including figure files
\usepackage{amsmath}	% Advanced maths commands
\usepackage{amssymb}	% Extra maths symbols
\usepackage{courier}
\usepackage{caption}
\usepackage{txfonts}
\title[{Smaller disc scale lengths in rich environments}]{Smaller stellar disc scale lengths in rich environments}

\author[M. L. Demers et al.]{
Melanie L. Demers,$^{1}$\thanks{E-mail: demerm2@mcmaster.ca}
Laura C. Parker,$^{1}$\thanks{E-mail: lparker@mcmaster.ca}
Ian D. Roberts$^{1}$
\\
$^{1}$Physics and Astronomy, McMaster University, 1280 Main Street West, Hamilton L8S 4L8, Canada\\
}

\date{Accepted XXX. Received YYY; in original form ZZZ}

\pubyear{2019}

\begin{document}
\label{firstpage}
\pagerange{\pageref{firstpage}--\pageref{lastpage}}
\maketitle

\defcitealias{Yang2007}{Y07}
\defcitealias{Simard2011}{S11}
\defcitealias{Mendel2014}{M14}

\begin{abstract}

We investigate the dependence of stellar disc scale lengths on environment for a sample of Sloan Digital Sky Survey Data Release 7 galaxies with published photometric bulge-disc decompositions. We compare disc scale lengths at fixed bulge mass for galaxies in an isolated field environment to galaxies in X-ray rich and X-ray poor groups. At low bulge mass, stellar disc scale lengths in X-ray rich groups are smaller compared to discs in both X-ray poor groups and in isolated field environments. This decrease in disc scale length is largely independent of halo mass, though shows some dependence on group-centric distance. We also find that stellar disc scale lengths are smaller in X-ray rich environments for a subset of star-forming galaxies and for galaxies of different morphological types. We note that disc scale lengths of low mass galaxies are known to have large systematic uncertainties, however we focus on differences between samples with the same measurement biases. Our results show that stellar disc scale lengths depend on X-ray brightness, a tracer of IGM density, suggesting a role for hydrodynamic processes such as ram-pressure stripping and/or starvation. 
\end{abstract}


%%%%%%%%%%%%%%%%%%%%%%%%%%%%%%%%%%%%%%%%%%%%%%%%%%

%%%%%%%%%%%%%%%%% BODY OF PAPER %%%%%%%%%%%%%%%%%%

\section{Introduction}

Galaxies in the local universe follow a largely bi-modal distribution in many observed properties \citep{Brinchmann2004, Baldry2006, Driver2006, Bamford2009, Wetzel2012}. In the star-formation rate (SFR) - stellar mass plane, these two populations of galaxies are separated into an actively star-forming ``blue cloud" and a quiescient ``red sequence" \citep[e.g.][]{Brinchmann2004}. Blue cloud galaxies are typically disc-dominated; whereas red sequence galaxies are often bulge-dominated. Previous work has suggested that galaxies evolve onto the red sequence through two distinct pathways: mass quenching and environmental quenching \citep[e.g.][]{Peng2010}.

Environmental quenching is often used to explain the morphology-density relation, whereby red bulge-dominated galaxies are more likely to be found in dense environments, and blue disc-dominated galaxies are more commonly found in underdense environments \citep{Oemler1974, Dressler1980, Lewis2002, Goto2003, Gomez2003, Balogh2004, Yang2007, Bamford2009}. Although there are some exceptions to these trends - such as blue ellipticals \citep{Schawinski2014} and red spirals \citep{Masters2010, Evans2018} - the general bi-modality of galaxy colour and morphology holds true. 

There are a host of environmental mechanisms invoked to explain the morphology-density relation. Galaxy mergers (both major and minor) have been proposed to explain the build-up of elliptical galaxies and the formation of bulges in the centres of spiral galaxies with time \citep[e.g.][]{ToomreToomre1972, MakinoHut1997, Angulo2009, Wetzel2009a, Wetzel2009b, White2010, Cohn2012}. Other galaxy-galaxy interactions such as harassment and tidal encounters can also remove the gaseous and stellar discs of galaxies \citep{Moore1996, Moore1998, Boselli2006, Cortese2006}. These effects are more likely to occur in dense environments, although major mergers are rare in clusters due to high relative velocities \citep[e.g.][]{Conselice2009}.

Galaxies also interact with the hot gas permeating group and cluster environments known as the intra-group or intra-cluster medium (IGM hereafter). Starvation (or strangulation), a process that prevents the cooling and accretion of hot halo gas onto galactic discs, or strips away hot halo gas via the IGM, is known to quench star-formation and can convert late-type spirals into early-type morphologies over long time scales \citep{Larson1980, Bekki2002, vandenBosch2008}. Based on observational results and simulations, it has been shown that starvation quenches star-formation on a timescale of $\gtrsim 1\,\mathrm{Gyr}$ \citep{OkamotoNagashima2003, Fujita2004, Tonnesen2007, Moran2007, KawataMulchaey2008}. Although, after star-formation stops, the gradual process of starvation preserves spiral structure for over $\sim 1\,\mathrm{Gyr}$ before changes in morphology begin to take place \citep{Bekki2002, Boselli2006, Moran2006}. Since starvation is a slow-acting process, many observational studies at local redshift conclude that starvation only has a small affect on morphology \citep{vandenBosch2008, Schawinski2014}.

Ram-pressure stripping \citep[e.g.][]{GunnGott1972} has also been observed to quench star-formation by removing atomic hydrogen gas from disc edges \citep[e.g.][]{Yagi2007, Zhang2013}. In general, atomic hydrogen discs are more extended than stellar discs, and the stellar component is more tightly bound \citep[e.g.][]{Kennicutt1989}. This implies that atomic hydrogen is more readily stripped than stars via processes such as ram-pressure stripping \citep{Abadi1999, Joshi2017}. This is consistent with observations showing that disc galaxies have reduced ratios of gas/stellar scale lengths in clusters as compared to field environments \citep{Bosch2013}. Models and simulations have shown that ram-pressure stripping can remove the entire gaseous component of discs in $\sim 100\,\mathrm{Myr}$ in high density environments \citep{FujitaNagashima1999, Quilis2000, RoedigerHensler2005, Kapferer2009}. Additionally, toy modelling of ram-pressure stripping studying the response of the stellar disc to the ram-pressure drag force on the gas disc has also shown that the \textit{stellar disc} can be displaced by several kiloparsecs in the direction of the ram-pressure wind \citep{Smith2012}.

Environmental quenching mechanisms caused by the IGM, such as starvation and ram-pressure stripping, can possibly be traced by studying the X-ray luminosity of groups. For example, \cite{ZabludoffMulchaey1998} find that X-ray bright groups (with the densest IGMs) have a higher fraction of early-type galaxies than X-ray undetected groups. However, not all studies find a dependence of morphology on X-ray environment. For example, \cite{Poggianti2016} find candidates for galaxies undergoing ram-pressure stripping in clusters and groups and these galaxies are found at all group-centric radii and their number does not correlate with X-ray host luminosity $L_{x}$.

The mass-size relation \citep[e.g.][]{Shen2003} is another useful tool for interpreting the growth of galaxies as a function of both redshift and of environment. Many studies have found that galaxy effective radii grow from redshifts 2-3 until today  \citep{Daddi2005, Trujillo2006, Longhetti2007, Trujillo2007, Cimatti2008, vanDokkum2008}. To focus on environmental influences, previous studies have compared the mass-size relation at fixed redshift in different environments. At low \citep{Pohlen2006, Maltby2010, Maltby2015} and high \citep{Rettura2010, Valentinuzzi2010} redshift, studies have shown that the mass-size relation shows no environmental dependence for early-type and lenticular galaxies. However, \cite{Maltby2010} show that at low redshift, late-type spiral galaxies with $\rm \log(M_{\star}/M_{\odot}) < 10$ do depend on their environments, with effective radii that are $\sim 15-20$\% larger in the field than in clusters. They conclude that environment can possibly play a role in reducing the sizes of the discs of low-mass spirals \citep[in agreement with][]{Tasca2009}. Further, \citet{FernandezLorenzo2013} showed that the discs of more massive spirals also display environmental dependencies.

Taking one step further, we can explore the effects of environment on bulges and discs separately. In order to study how the structural components of galaxies evolve with environment \citet{Driver2006} suggest modelling galaxies as two-component bulge + disc systems. There is an extensive body of literature that examines bulge scaling relations \citep[e.g][]{Magorrian1998, MarconiHunt2003, HaringRix2004, Graham2012, Graham2014, Bluck2014}, and properties of spiral galaxies such as luminosity, rotational velocity, and disc size also obey known scaling relations \citep[e.g.][]{Courteau2007, Dutton2011}. However, there is a relative lack of studies addressing the dependence of disc scaling relations on environment. If the bulges of low-mass spiral galaxies have not experienced growth in stellar mass after $z \sim 1$ \citep{vanDokkum2013}, then this implies galaxy discs continued to change with environment over cosmic time, and could drive overall changes in morphology \citep{Hudson2010}.

We analyze disc scale length at fixed bulge mass in isolated field and group environments, using X-ray brightness as a proxy for the IGM density in groups. In \textsection \ref{sec: 2}, we describe our sample selection and data analysis. In \textsection \ref{sec: 3}, we present the disc scale length versus bulge mass relation for our sample. We discuss the implications of these results in \textsection \ref{sec: 4} and present our conclusions in \textsection \ref{sec: 5}. This work assumes a $\Lambda$CDM cosmology with $\Omega_{m} = 0.3$, $\Omega_{\Lambda} = 0.7$, and $h = 0.7$.

\section{Data}
\label{sec: 2}

To examine the dependencies of galaxy discs on environment, we make use of a wealth of data from several public catalogues consisting of Sloan Digital Sky Survey Data Release 7 \citep[SDSS-DR7;][]{Abazajian2009} galaxies and groups.

\subsection{Group properties} 
\label{sec:catalogues}

\subsubsection{Group environment}

We use a large sample of groups identified in SDSS-DR7 by \cite{Yang2007}, hereafter \citetalias{Yang2007}. The group finder developed by \cite{Yang2005, Yang2007} applies a modified Friends of Friends (FOF) algorithm \citep[e.g][]{HuchraGeller1982, Davis1985} to the sample, and assigns galaxies to tentative groups based on short linking lengths in redshift-space. First, the FOF algorithm identifies tentative luminosity-weighted group centres. Next, \citetalias{Yang2007} calculate the characteristic luminosity of all assigned group members, such that the flux limit of the SDSS survey at different redshifts is taken into account. Subsequently, the group finder determines the halo mass of each tentative group, based on an assumed mass-to-light ratio and the luminosity of the group members. Using the halo mass, the halo radius and velocity dispersion are calculated for each group. Based on these halo parameters, \citetalias{Yang2007} adds or removes galaxies via their phase-space information to refine group membership. They repeat this process iteratively until the assigned group membership remains unchanged (see \citetalias{Yang2007} for a complete description).

We calculate group-centric distances via the angular separation of each galaxy from the luminosity weighted centre of its host group following \cite{Hogg1999}. We normalize group-centric distances by the host group's virial radius $R_{200m}$, calculated in \cite{TinkerChen2008}:

\begin{equation} \label{(eqn:1)}
R_{200m} = \left(\frac{3M_{halo}}{4\pi \cdot 200 \overline{\rho_{m}}}\right)^{1/3}
\end{equation}

\noindent
where $M_{halo}$ is the group halo mass given in \citetalias{Yang2007}, $\ \overline{\rho_{m}} = \Omega_{m,0} \ \rho_{c,0} \ (1+z_{g})^{3} $ is the average background matter density of the universe, and $z_{g}$ is the redshift of the luminosity-weighted group centre. Using our assumed cosmology, this can be simplified to:

\begin{equation} \label{(eqn:2)}
R_{200m} = 1.13 h^{-1}\left(\frac{M_{halo}}{10^{14}h^{-1}\mathrm{M_{\odot}}}\right)^{1/3}(1+z_{g})^{-1} 
\end{equation}
\noindent

\subsubsection{X-ray properties}
\label{sec: 2.1.2}
We use group X-ray luminosities determined by \cite{Wang2014}, who estimated the X-ray luminosities for ~65,000 optically selected galaxy groups (and clusters) in the SDSS using X-ray data from the ROSAT All Sky Survey \citep[RASS;][]{Voges1999}. \cite{Wang2014} use an Optical to X-ray (OTX) code \citep[see][]{Shen2008} to measure group X-ray luminosities. The OTX code begins with an optically identified group with $\rm \log(M_{halo}/M_{\odot}) \gtrsim 13 $ and identifies the most-massive galaxies (MMGs), keeping up to 4 MMGs for each group. Using a maximum likelihood algorithm, the RASS sources that are associated with the MMGs are identified, while masking out contaminant sources such as quasi-stellar objects and stars. The algorithm determines the X-ray background for each group by measuring the number of counts enclosed in a 6 arcminute wide annulus, whose inner ring corresponds to $R_{200m}$, that is centered on the X-ray centre. The X-ray background is subtracted off, and the cumulative source count rate profile as a function of radius is calculated for each group. Integrating the source count rate profile within $R_{x} = 0.5\, R_{200m}$ results in an X-ray luminosity $L_{x}$ for each group. 

\cite{Wang2014} applies the OTX code to $\sim65,000$ optical groups, and after background subtraction $\sim34,500$ groups have signal-to-noise $S/N > 1$ in the X-ray. If we restrict this to galaxies with redshifts less than 0.1, so that our sample is relatively complete to $\rm \log(M_{\star}/M_{\odot}) \sim 9$, the group sample is reduced by $\sim 50$\%. We also restrict halo masses (obtained from \citetalias{Yang2007}) to be within  $\rm 13 \leq \log(M_{halo}/M_{\odot}) \leq 15 $, which results in a sample of $\sim 4000$ groups \citep[as in][]{Roberts2016}. The machine readable versions of our catalogues with galaxies that reside in X-ray detected groups out to $z \leq 0.1$ is described in Appendix~\ref{appendix: a}, Table~\ref{tab: tab1}.

\begin{figure*}
	\centering
	\begin{minipage}[b]{.45\textwidth}
		\includegraphics[width=1\textwidth]{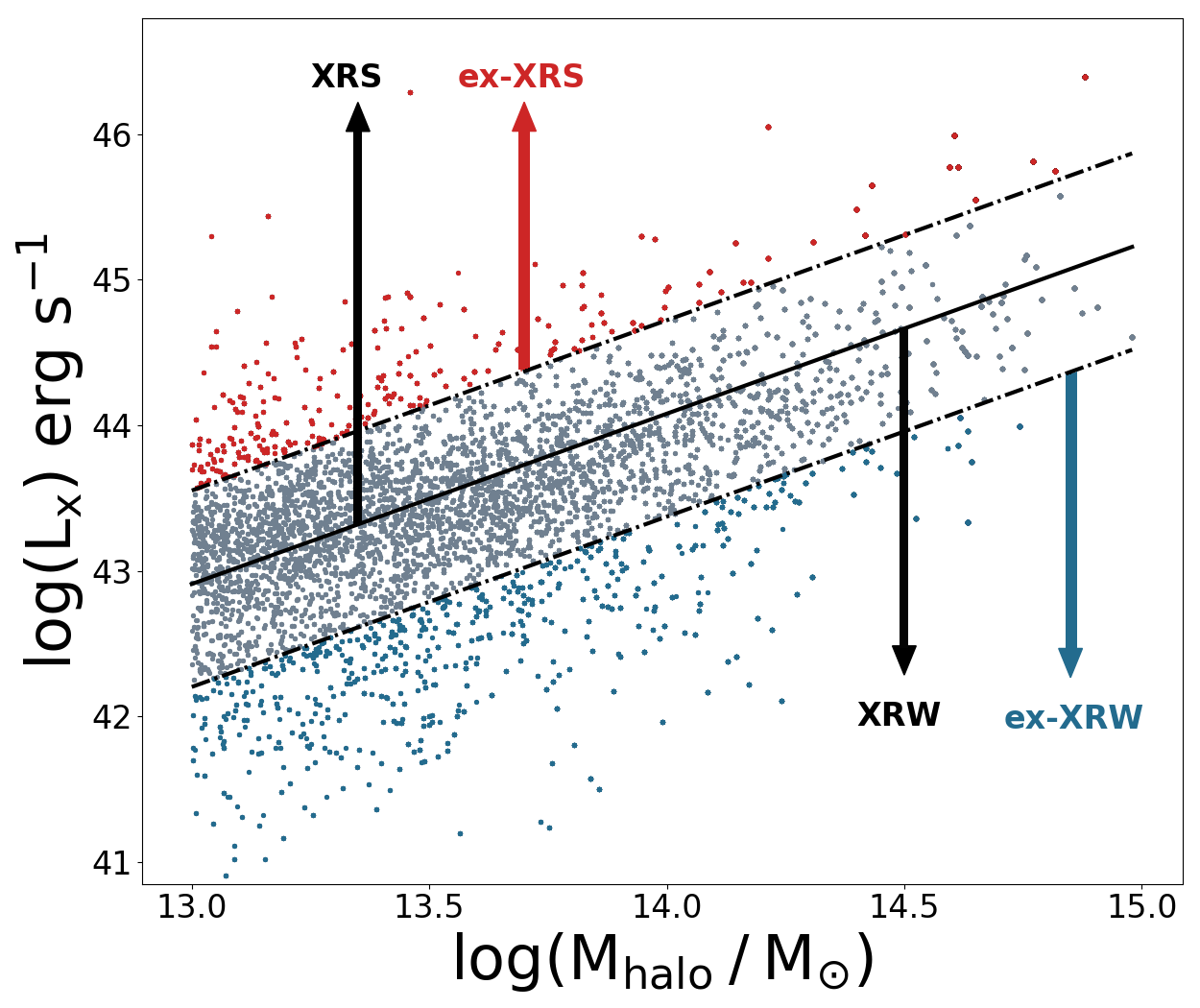}
		\caption{Galaxy group samples in the $L_{x} - M_{halo}$ plane. Groups above and below the solid line of best fit delineate the X-ray strong (XRS) and X-ray weak (XRW) group samples. The top (red) and bottom (blue) deciles of the dataset that lie above and below the dashed lines are referred to as the extremely X-ray strong (ex-XRS) and extremely X-ray weak (ex-XRW) group samples.}
		\label{fig1}
	\end{minipage}%
	\hspace{1cm}
	\begin{minipage}[b]{.45\textwidth}
		\includegraphics[width=1\textwidth]{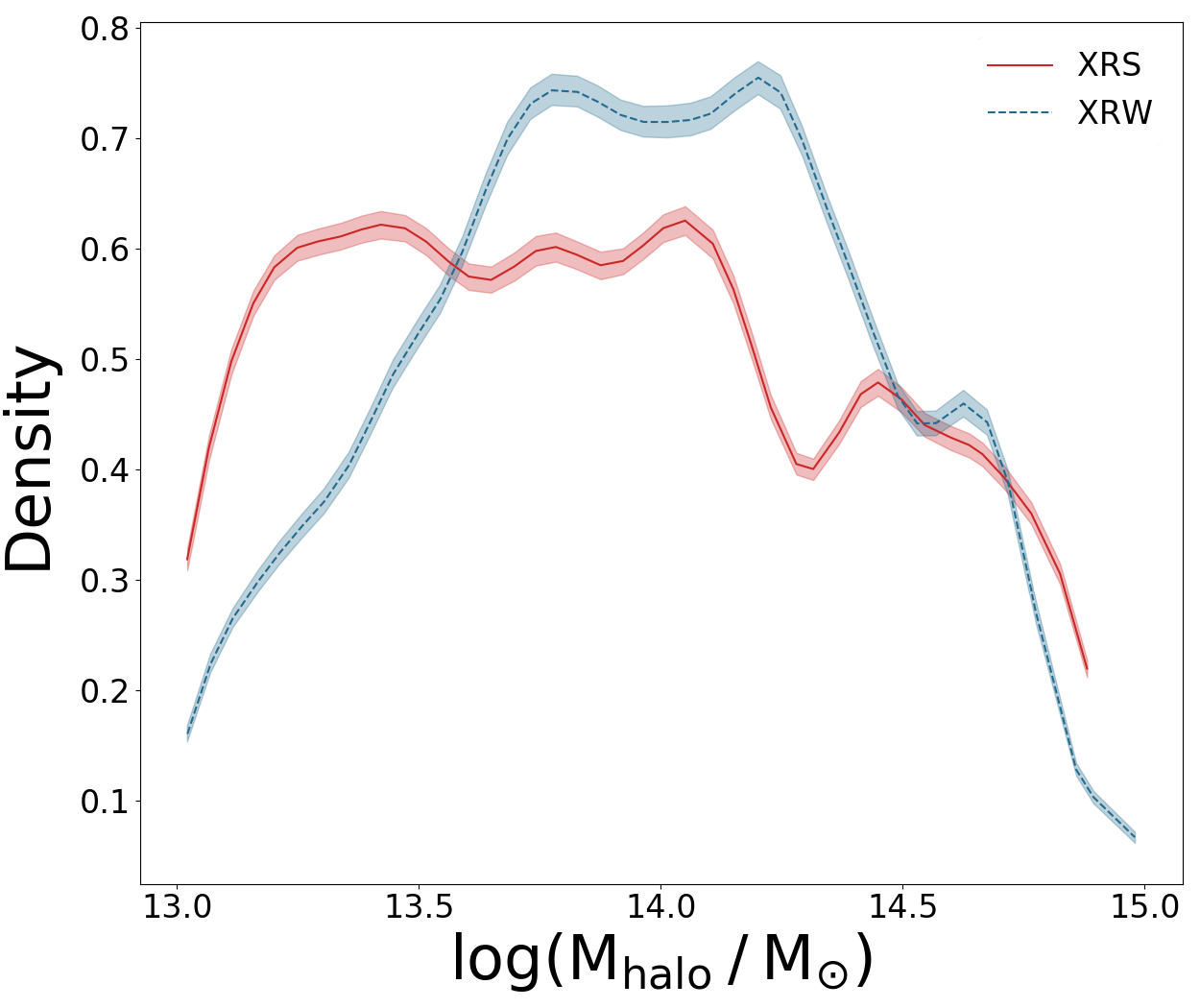}
		\caption{Smoothed density distributions of halo masses in the XRS (red) and XRW (blue) group samples. Shaded regions correspond to the 16th and 84th confidence levels from 1000 bootstrap resamplings.\newline\newline\newline}
		\label{fig2}
	\end{minipage}
\end{figure*}

To study the effects of the group environment\footnote{Typically clusters are defined to have halo masses of $\rm \log(M_{halo}/M_{\odot}) \gtrsim 14$, but throughout our work we refer to all systems with $\rm \log(M_{halo}/M_{\odot}) > 13$ as groups.} on galaxies, we divide this sample into two approximately equal subsamples, based on whether a given group lies above or below the line of best fit in the $L_{x} - M_{halo}$ plane \citep[see][]{Wang2014, Roberts2016}. The galaxies in groups above this line are hereafter referred to as the X-ray strong (XRS) sample, and the galaxies that are members of groups below this line are referred to as the X-ray weak (XRW) sample (see Fig.~\ref{fig1}). The halo mass distribution for the XRS and XRW samples our shown in Fig.~\ref{fig2}.

We also target the most extreme ends of our dataset in the  $L_{x} - M_{halo}$ plane (see Fig.~\ref{fig1}), by shifting the y-intercept of the line of best fit upwards (or downwards), until the top (or bottom) decile of the dataset is isolated. We will subsequently refer to these extreme ends of our sample as extremely X-ray strong (ex-XRS) and extremely X-ray weak (ex-XRW) (see \textsection \ref{sec: 4.2}).

\subsection{Galaxy properties}

\subsubsection{Structural parameters}

We use bulge and disc decompositions of SDSS galaxies from \cite{Simard2011}, hereafter \citetalias{Simard2011}. \citetalias{Simard2011} uses the \texttt{GIM2D} software package \citep[see][]{Simard2002} to fit the two-dimensional surface brightness profiles of SDSS galaxies in the \textit{g} and \textit{r} bandpasses. There are three parametric models used for these fits: i) a fixed central bulge with a de Vaucouleurs profile $n_{b} = 4$ and a free disc ii) a free bulge and a free disc iii) a pure single-component S\'ersic fit. In this work we use the structural parameters from a fixed de Vaucouleurs bulge and free disc (table 1 in \citetalias{Simard2011}). There are many structural parameters measured, but we specifically use the half-light radius and exponential disc scale length derived from the following expressions \citep{Sersic1968}:

\begin{equation} \label{(eqn: 3)}
\Sigma(r) = \Sigma_{e}\exp{(-k[(r/R_{e})^{1/n} -1] )}
\end{equation}

\noindent
where $\Sigma_{e}$ is the surface brightness of the galaxy at effective radius $R_{e}$, $n$ is the S\'ersic index, and  $ k = 1.9992n - 0.3271$ \citep{Capaccioli1989}. Elliptical galaxies and the classical bulges of spiral galaxies are well-described by a de Vaucouleurs profile with $n=4$. Whereas, galaxy discs have been shown to be well-fit by an $n=1$ surface brightness profile (\citealt{Andredakis1995, deJong1996}; \citetalias{Simard2011}), such that:

\begin{equation} \label{(eqn: 4)}
\Sigma(r) = \Sigma_{0}\exp{(r/R_{d})}
\end{equation}

\noindent
where $\Sigma_{0}$ is the face-on surface brightness of the galaxy, and $R_{d}$ is the exponential disc scale length.

\subsubsection{Bulge and disc properties}

We obtain the total stellar, bulge, and disc masses from \cite{Mendel2014}. The masses derived in the \cite{Mendel2014} catalogue are based on an extension of \citetalias{Simard2011}, with additional S\'ersic decompositions of SDSS galaxies in the \textit{u}, \textit{i} and \textit{z} bandpasses. To estimate stellar masses, each galaxy's spectral energy distribution (SED) is compared to a library of synthetic stellar populations, generated by the stellar population synthesis (SPS) code developed and calibrated by \cite{Conroy2009, Conroy2010first} and \cite{Conroy2010second}. Each SPS library is constructed to incorporate a range of stellar ages, metallicities, star-formation histories, and dust parameters. There are two SPS libraries generated: i) one that accounts for internal dust attenuation in the SPS grid, following the extinction law of \citep{Calzetti2000} ii) the other fixes \textit{E(B-V)} = 0, and assumes a dust-free model. A Bayesian approach is used to carry out the SED fits for a given SPS library, where a likelihood function describing the flux of each object in the \textit{u}, \textit{g}, \text{r}, \textit{i}, \text{z} bands is defined, and the priors are defined via the construction of the SPS grid. We use the dust-corrected stellar, bulge and disc mass estimates, but we find no differences in our results when using dust-free estimates.

\subsection{Final sample}
\label{sec: 2.3}

\subsubsection{Group galaxies}
\label{sec: 2.3.1}

\begin{figure}
	\includegraphics[width=\columnwidth]{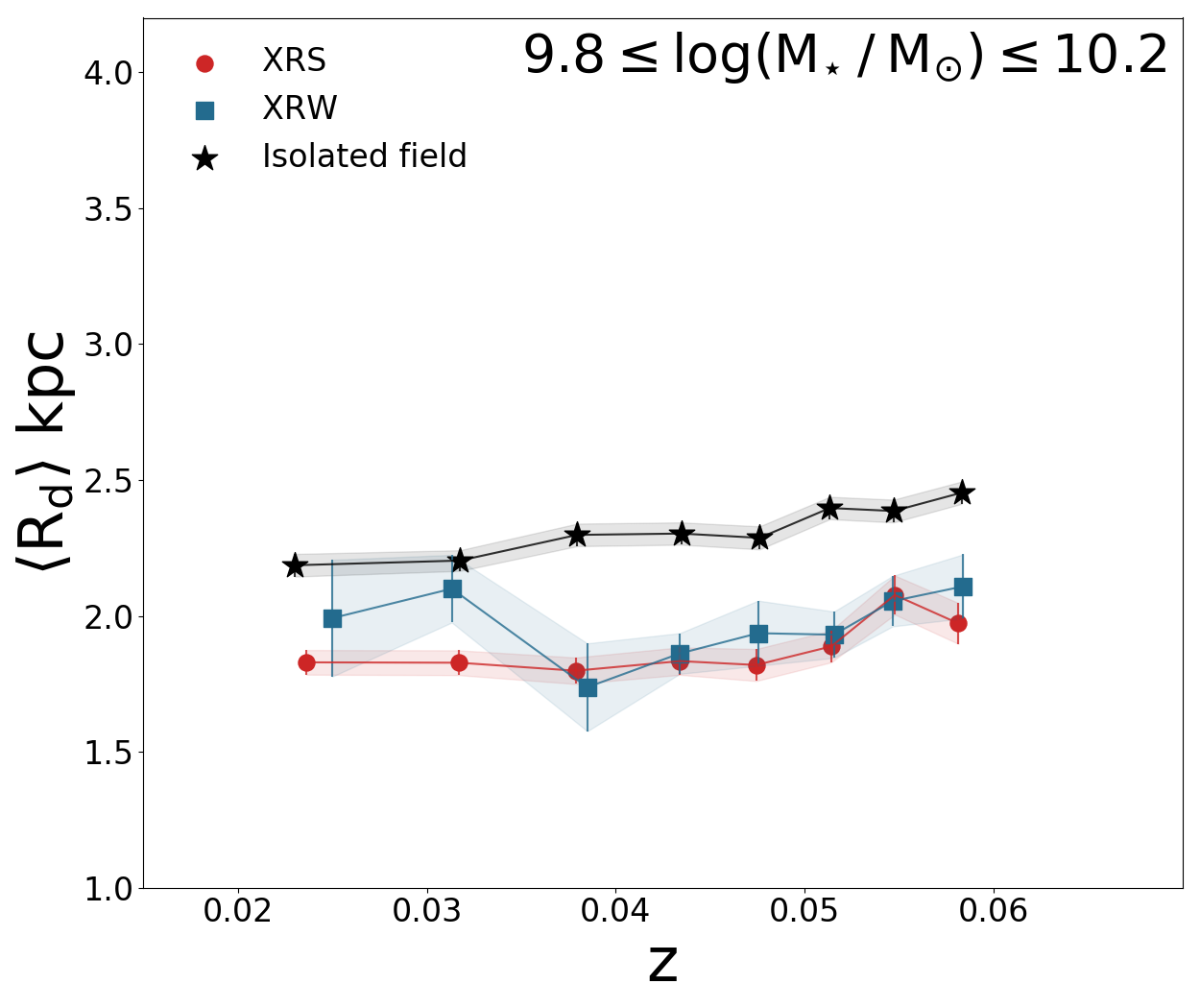}
	\caption{$V_{max}$ weighted exponential disc scale length versus redshift for X-ray strong (XRS), X-ray weak (XRW), and isolated field samples with $9.8 \leq \log(M_{\star}/M_{\odot}) \leq 10.2$. Error bars and shaded regions correspond to the standard error of the mean.}
	
	\label{fig3}
\end{figure}

Given the relatively poor image quality of the SDSS, we further restrict our sample to ensure we have reliable measurements of structural parameters. First, we determine an appropriate redshift cut by examining the dependence of disc scale length on redshift for a narrow range in stellar mass. In Fig.~\ref{fig3}, we examine the exponential disc length versus redshift for X-ray (XRS), X-ray weak (XRW) and isolated field galaxies at fixed stellar mass for $9.8 \leq \log(M_{\star}/M_{\odot}) \leq 10.2$. At fixed stellar mass, we see no strong dependence of exponential disc scale length on redshift at $z \leq 0.06$. Similar redshift cuts have been suggested in other SDSS studies of galaxy structural properties \citep[e.g.][]{Meert2015}.

\citetalias{Simard2011} also define a parameter $P_{pS}$ as the probability that a bulge + disc model is not required compared to a single component S\'ersic model. To distinguish which galaxies are best represented by genuine bulge + disc systems, it is recommended to apply a $P_{pS} \leq 0.32$ cut which we use throughout our analysis. We also limit our sample to include galaxies with bulge effective radii $R_{e} \geq 0.1"$, to eliminate point-like bulges. The final group sample we use in our work has 4,452 galaxies in 866 host groups, where each galaxy has measured structural parameters from \citetalias{Simard2011} and \cite{Mendel2014}. We note that these additional cuts result in our XRW sample being significantly smaller than our XRS sample (defined in \textsection \ref{sec: 2.1.2}). 

Since our final samples are not volume-limited, we correct for the Malmquist bias by applying $V_{max}$ weights (given by \citetalias{Simard2011}) to our results, where $V_{max}$ is the volume of the sample corresponding to the largest distance that a galaxy with a measurement of absolute magnitude can be observed in order to have an apparent magnitude equal to the magnitude limit of the sample. We apply a $1/V_{max}$ weighted average to our galaxy measurements as follows:

\begin{equation} \label{(eqn: vmax)}
\frac{\sum_{i = 1}^{N}   x_{i} \cdot 1/V_{max_i}}{\sum_{i = 1}^{N} 1/V_{max_i}}
\end{equation}

\noindent
where $x_{i}$ is an observed galaxy property $x$ at $i$, and $N$ is the total sample size. We note that due to the small sample range in redshift of our sample ($z \leq 0.06$) our key results are insensitive to $V_{max}$ weighting.
	
The group galaxy sample spans a redshift range of $0.01 \leq z  \leq 0.06$, stellar mass range of $\rm 9 \lesssim \log(M_{\star}/M_{\odot}) \lesssim 11.4$ and halo mass range of $\rm 13 \lesssim \log(M_{halo}/M_{\odot}) \lesssim 15$.

\subsubsection{Isolated field galaxies}
\label{sec: 2.3.2}
We compare our group sample to a field population of 46,647 galaxies at $z \leq 0.06$ with measured structural parameters, using an isolated field catalogue constructed by \citet{Roberts2017} based on \citetalias{Yang2007}. \citet{Roberts2017} apply isolation criteria to single member group galaxies from \citetalias{Yang2007} such that: the projected group-centric distance of each galaxy is greater than 3 virial radii; each galaxy has a line-of-sight velocity greater than $1.5\, \sigma$ from the group centre, where $\sigma$ is the velocity dispersion of the group centre (calculated from Eqn. 6 in \citetalias{Yang2007}); and each galaxy is separated from its nearest bright neighbour by a minimum of $ 1\,\mathrm{Mpc}$ in projected distance and $1000\,\mathrm{kms^{-1}}$ in line-of-sight velocity. In this work, bright neighbours are defined as galaxies which are brighter than the SDSS survey $r$ band magnitude limit at $ z=0.06$, $M_{r,lim}=-19.4$. Galaxies that are within $1\,\mathrm{Mpc}$ of SDSS survey edges are removed, as well as galaxies within $1000\,\mathrm{kms^{-1}}$ of $z=0.06$.

As we have done for our group samples, we also apply further cuts to the isolated field sample where $P_{pS} \leq 0.32$ and $R_{e} \geq 0.1"$ (as described in \textsection \ref{sec: 2.3.1}) which reduces our final sample size of isolated field galaxies to 15,234. We have included a machine readable table for the isolated field catalogue for future comparison to this work described in Appendix~\ref{appendix: a}, Table~\ref{tab: tab2}.

\section{Results}
\label{sec: 3}

\subsection{Disc scale lengths at fixed stellar mass}

We first examine disc scale length as a function of galaxy total stellar mass. In Fig.~\ref{fig5} we show that the exponential disc scale length increases with stellar mass \citep[in agreement with][]{Jaffe2018} in all of our samples. At stellar masses below $\sim 10^{10.5}\, \mathrm{M_{\odot}}$ the group galaxies (XRS and XRW) have smaller discs than the isolated field galaxies. At higher stellar masses, the disc scale lengths of the group galaxies match those of the isolated field sample.

Fig.~\ref{fig5} shows differences between galaxies in XRS and XRW group environments. We find that at low stellar mass stellar discs in XRS groups are smaller than discs in XRW groups. It is worth noting that exploring disc properties at fixed galaxy stellar mass may not be very sensitive to changes in the disc, since any stripping of stars in the disc would also lead to changes in total stellar mass.

\subsection{Disc scale lengths at fixed bulge mass}

To more directly account for possible changes in the disc with environment, we investigate how disc scale lengths change with environment at fixed bulge mass rather than at fixed stellar mass \citep[a similar approach has been used in recent works such as][]{LacknerGunn2013, Bluck2014, Lang2014}.

In Fig.~\ref{fig6}, we observe that the disc scale length increases with bulge mass in the group (XRS and XRW) and isolated field samples. At low bulge masses, disc scale lengths are largest in the isolated field, smaller in XRW group environments, and smallest in XRS environments. These trends are consistent with the trends at fixed stellar mass in Fig.~\ref{fig5}. The disc scale lengths of galaxies in the XRS sample are smaller than in the isolated field by $\sim 1\,\mathrm{kpc}$ at low bulge mass. We also find that the disc scale lengths of galaxies in XRS groups are systematically smaller than those of galaxies in XRW groups by $\sim 0.5\,\mathrm{kpc}$ at low bulge mass.

\subsection{Trends with halo mass and group-centric position}

\begin{figure*}
	\centering
	\begin{minipage}[b]{.45\textwidth}
		\includegraphics[width=1\textwidth]{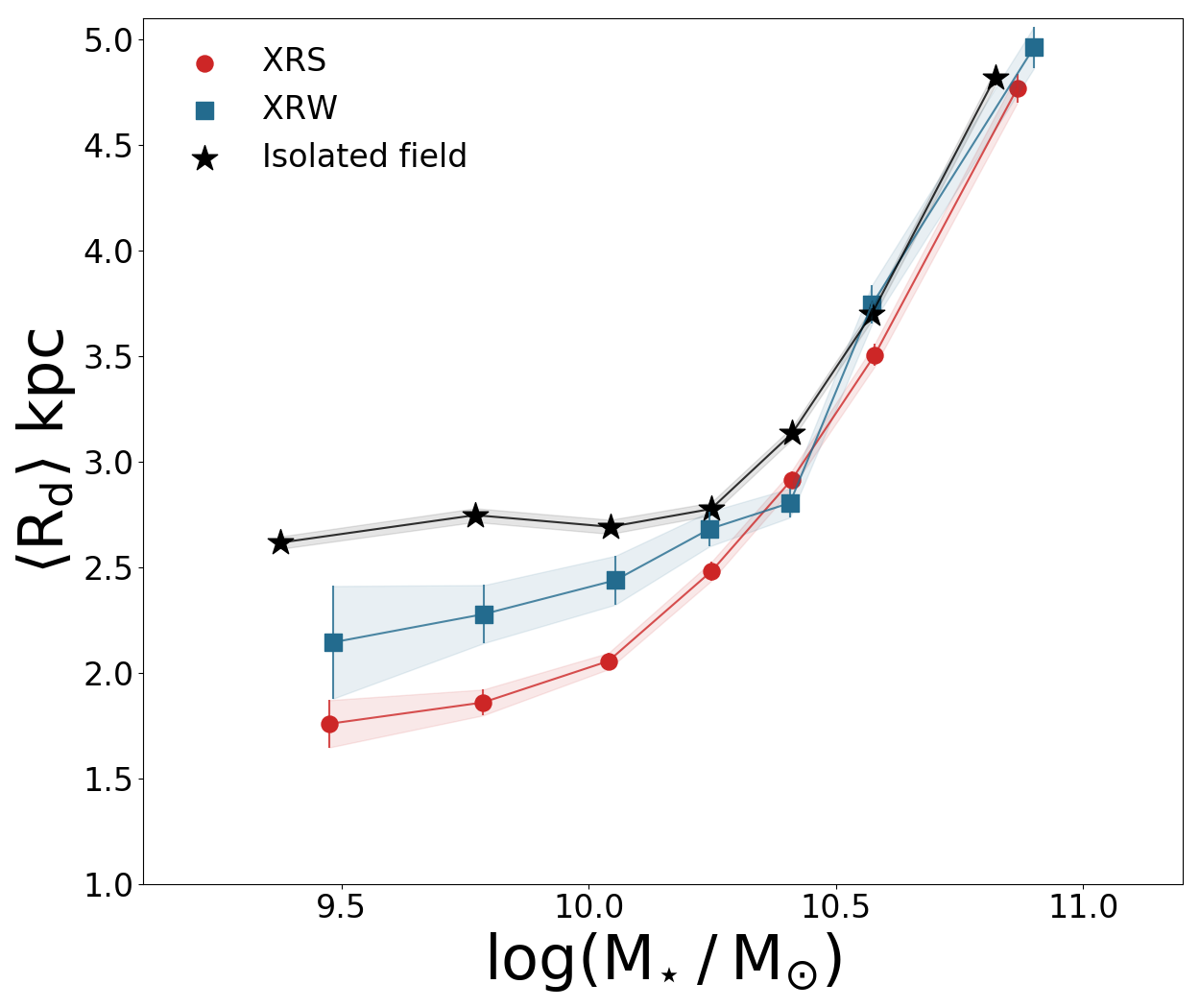}
		\caption{$V_{max}$ weighted exponential disc scale length versus stellar mass. Error bars and shaded regions correspond to the standard error of the mean.}
		
		\label{fig5}
	\end{minipage}%
	\hspace{1cm}
	\begin{minipage}[b]{.45\textwidth}
		\includegraphics[width=1\textwidth]{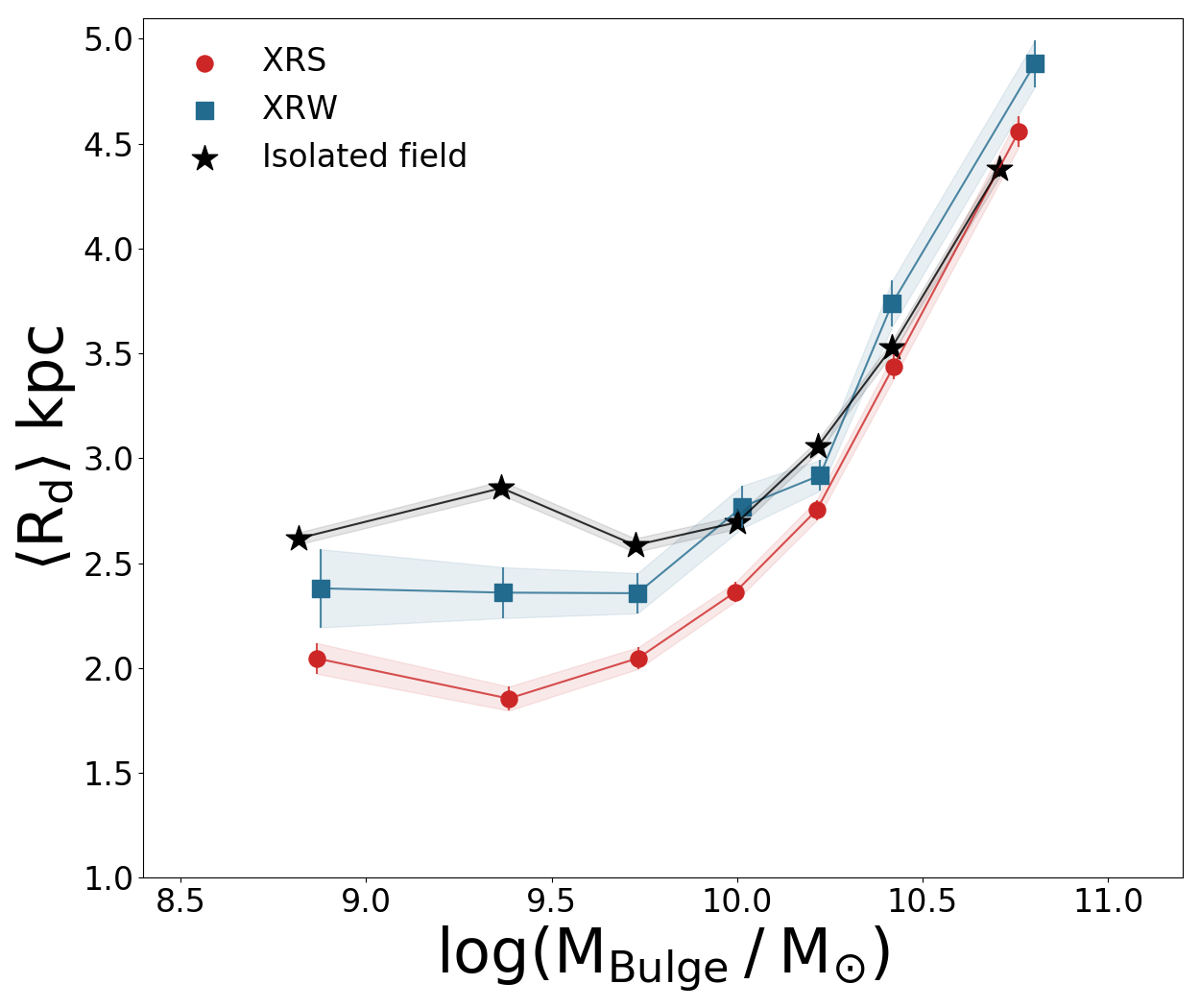}
		\caption{$V_{max}$ weighted exponential disc scale length versus bulge mass. Error bars and shaded regions correspond to the standard error of the mean.}
		\label{fig6}
	\end{minipage}
\end{figure*}

\begin{figure*}
	\includegraphics[width=\textwidth]{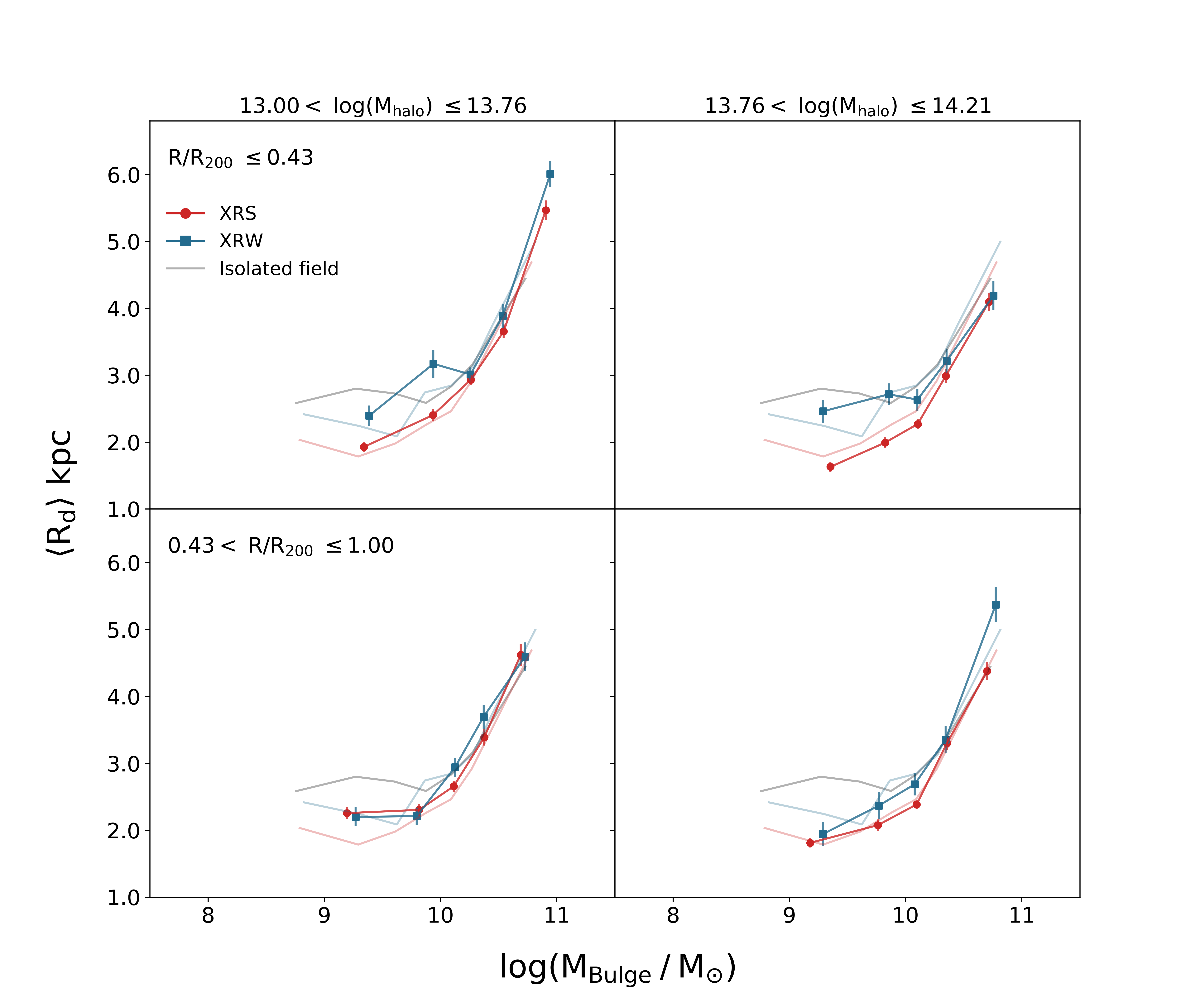}
	\caption{$V_{max}$ weighted exponential disc scale length versus bulge mass, in bins of increasing group-centric radius (top to bottom) and increasing halo mass (left to right). Average trends from Fig.~\ref{fig6} correspond to the faded lines. Bold lines correspond to the trend in each environmental panel, where the error bars correspond to standard error of the mean.}
	
	\label{fig7}
\end{figure*}

To further explore environmental dependencies, in Fig.~\ref{fig7} we show disc scale length versus bulge mass in bins of group-centric radius (normalized by $R_{200m}$) and halo mass. We note that the X-ray brightness of our XRS and XRW groups was previously defined at fixed halo mass in \textsection \ref{sec: 2.3.1}, and in this section we bin further by halo mass within our defined XRS and XRW samples. In Fig.~\ref{fig7}, the faded lines correspond to the average trend for total XRS, XRW, and isolated field samples shown in Fig.~\ref{fig6}. Overlaid on the average trends, in each panel we show the data in a given bin of group-centric distance and halo mass. Group-centric distance increases from top to bottom row, and halo mass increases from left to right. For example, the top left panel shows XRS and XRW data for small group-centric distances and low halo masses, overlaid on our average trends. Looking at all 4 panels, we find that the disc scale lengths in our XRW and XRS samples are largely independent of halo mass, though they do show a dependence on group-centric position. The difference in disc scale length between galaxies in XRW and XRS groups is largely coming from galaxies located close to their group centres.

\section{Discussion}
\label{sec: 4}

\subsection{Challenges with galaxy size measurements}
\label{sec: 4.1}

Reliable galaxy size measurements are notoriously challenging to obtain due to a variety of systematic effects. \cite{Mosleh2013} show that using a single component S\'ersic model (as opposed to a two component or non-parametric model) yields systematic overestimates of galaxy sizes at local redshifts, possibly due to substructure being missed in well-resolved galaxies. Furthermore, for early-type galaxies that are compact, the point-spread function (PSF) becomes a significant fraction of galaxy sizes with $R_{e} < 1\,\mathrm{kpc}$. In late-type galaxies, the edges of discs that have low-surface brightness components are also difficult to detect, due to low signal-to-noise. Similarly, \citet{Simard2002} state that bulge + disc structural parameters are most affected by systematic errors from sky background level determinations. If the sky level is underestimated, disc component sizes are overestimated due to positive sky residuals. Additionally, crowding errors and fibre errors in the SDSS redden galaxy colours, and this affects size determinations from surface brightness profile fits (see \citetalias{Simard2011} for their treatment of these errors). To distinguish the validity of applying bulge-disc decompositions, versus pure S\'ersic models to SDSS galaxies, \citetalias{Simard2011} provide a parameter $P_{pS}$ constructed using the F-statistic. $P_{pS}$ is defined as the probability that a bulge + disc model is not required compared to a single component S\'ersic model, and as described in \textsection \ref{sec: 2.3} we have applied this cut to all results shown in this work. 

The presence of dust in discs \citep[e.g.][]{Driver2007} can also change the intrinsic disc size-luminosity and colour-luminosity relations - although there is no simple prescription for applying empirical dust corrections to separate structural components of galaxies. Selecting face-on discs can mitigate the effects of dust; however, since our galaxies in XRS, XRW and isolated field samples have similar distributions in inclination, we did not apply dust corrections to our data. Overall, despite the challenges with determining individual galaxy sizes, we emphasize that this work is a relative comparison between three large samples with similar systematics. The relative differences between disc properties at fixed bulge mass in our samples are robust despite these uncertainties.

\subsection{Disc scale length at fixed stellar mass from Meert 2015}

We match our samples of group and isolated field galaxies from \textsection \ref{sec: 2.3} (with $z$, $M_{\star}$ and $P_{pS}$ cuts applied) to the \citet{Meert2015} catalogue of galaxies
fit with a de Vaucouleurs bulge and exponential disc model. A detailed description of the \citetalias{Simard2011}-\citet{Meert2015} matched sample is described in Appendix ~\ref{appendix: b}. Over the redshift range considered here ($0.01 \leq z  \leq 0.06$), we find that \citetalias{Simard2011} and \citet{Meert2015} disc scale lengths are well correlated (see Fig.~\ref{fig11}), with, for example, a value of RMS scatter of $0.275\, \mathrm{kpc}$ at $2\, \mathrm{kpc}$.

\begin{figure*}
	\centering
	\begin{minipage}[b]{.45\textwidth}
	\includegraphics[width=\columnwidth]{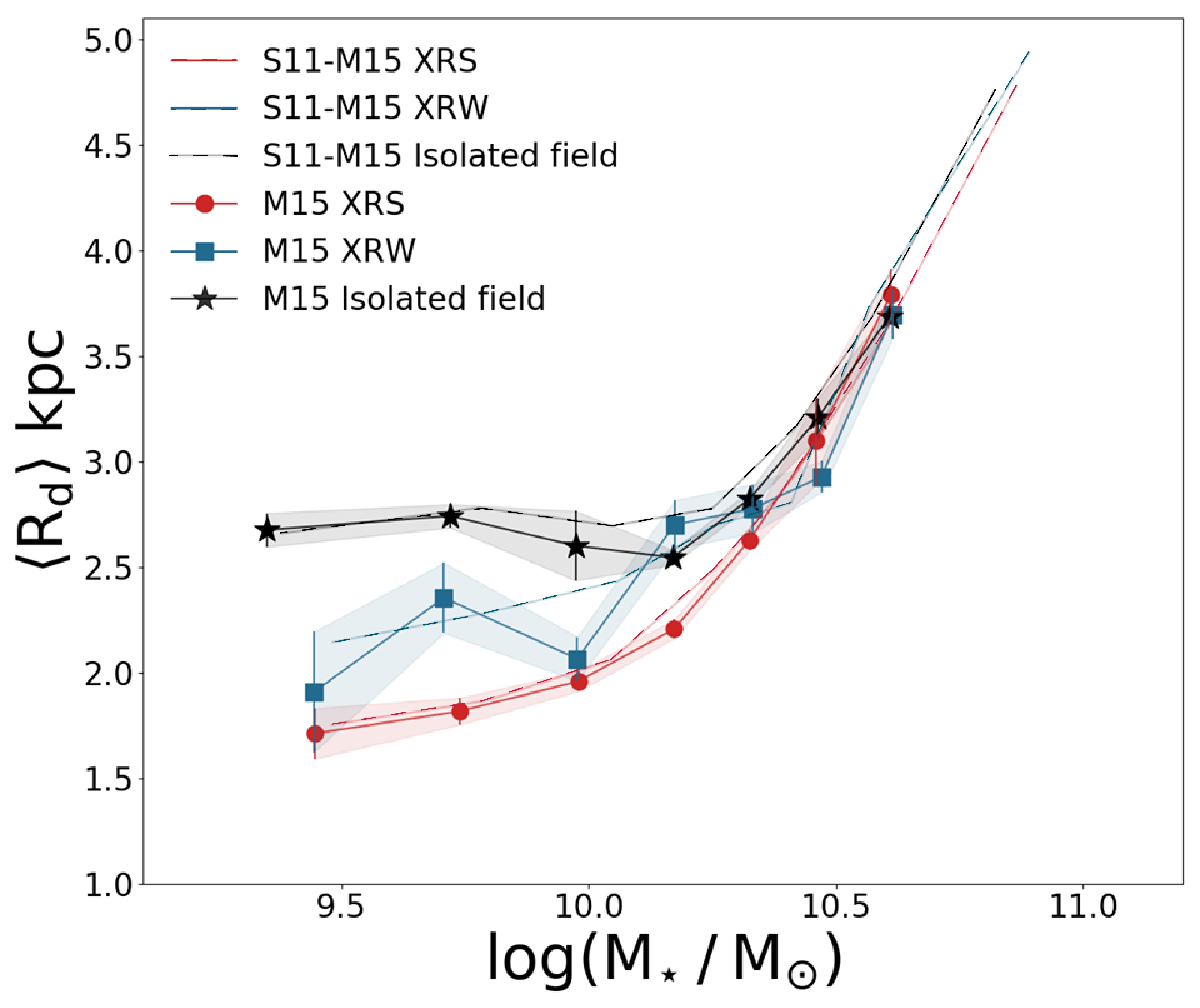}
	\caption{$V_{max}$ weighted exponential disc scale length versus stellar mass for \citet{Meert2015} $R_{d}$ values from \citetalias{Simard2011}-\citet{Meert2015} matched sample in Fig.~\ref{fig11} shown in bold lines for XRS, XRW, and isolated field samples, overlaid on \citetalias{Simard2011} XRS, XRW, and isolated field trends from Fig.~\ref{fig5} in faded lines.}
		\label{fig4}
	\end{minipage}%
	\hspace{1cm}
	\begin{minipage}[b]{.45\textwidth}
	\includegraphics[width=1\textwidth]{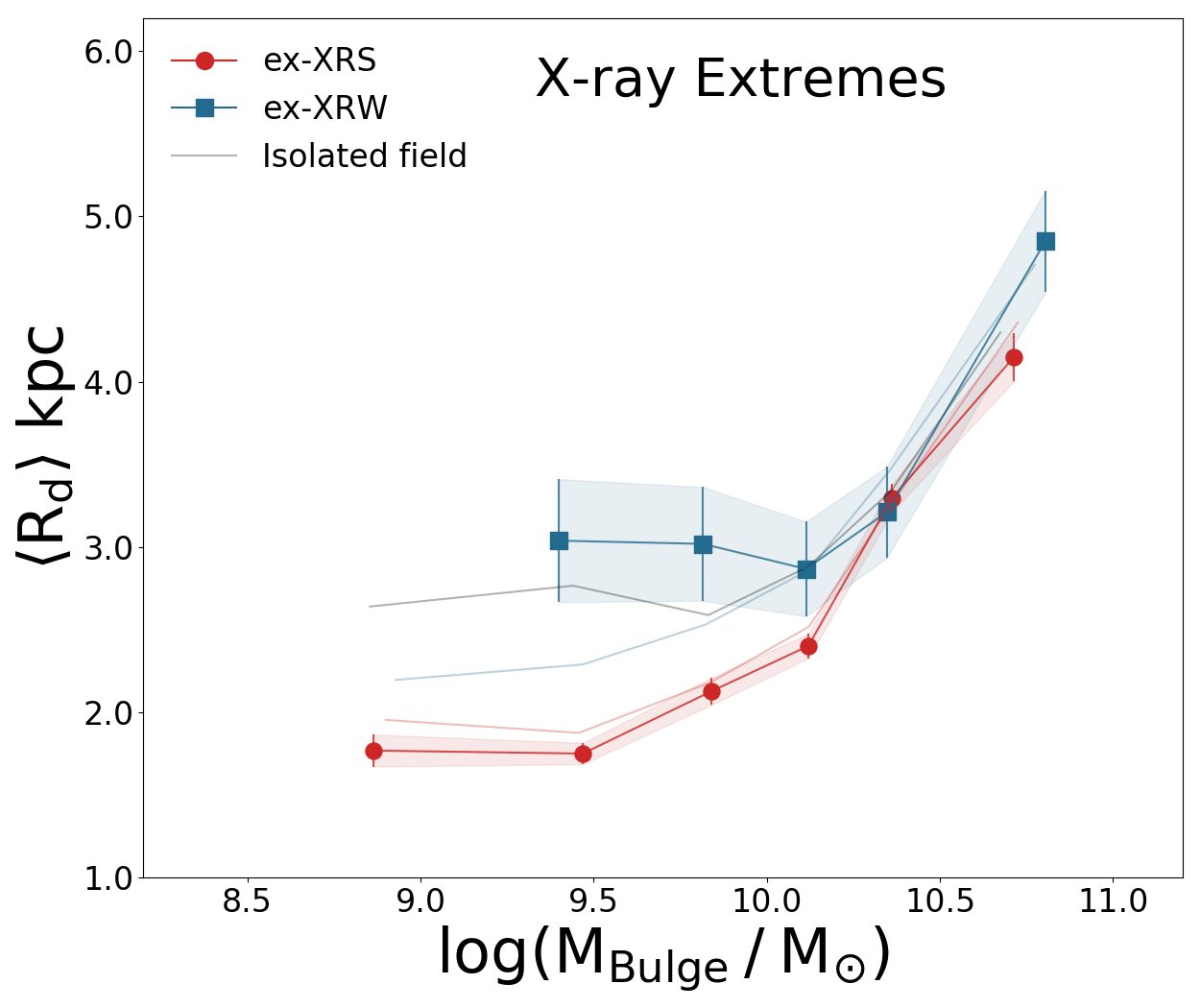}
	\caption{$V_{max}$ weighted exponential disc scale length versus bulge mass. Bold lines correspond to galaxies in extremely XRS and extremely XRW environments. Faded lines correspond to average trends for entire sample in Fig.~\ref{fig6}. Error bars and shaded regions correspond to the standard error of the mean.}
	\label{fig8}
\end{minipage}
\end{figure*}

In Fig.~\ref{fig4}, we show the exponential disc scale length from \citet{Meert2015} versus stellar mass in bold lines, overlaid on the trends found in Fig.~\ref{fig5} in faded lines. We show trends for disc scale length versus stellar mass, rather than bulge mass, since there are no bulge mass measurements provided in \citet{Meert2015}. We find that at low stellar mass, the disc scale lengths in \citet{Meert2015} display a systematic difference between XRS, XRW and field environments. We show that the \citet{Meert2015} results in Fig.~\ref{fig4} are in agreement with the \citetalias{Simard2011} results shown in Fig.~\ref{fig5}, and we note that the XRW disc scale lengths from \citet{Meert2015} have larger scatter. Our results are consistent whether we use \citetalias{Simard2011} or \citet{Meert2015} disc scale lengths; however, we choose to focus on the \citetalias{Simard2011} measurements, in combination with \citet{Mendel2014}, which additionally provide bulge mass measurements.

\subsection{X-ray extreme environments}
\label{sec: 4.2}

Our results in Fig.~\ref{fig6} show that galaxy disc scale length depends on group X-ray brightness. However, since the scatter in the X-ray luminosity for a given group is large \citep[of order 0.4 dex in][]{Wang2014}, we also look at our trends in disc scale length at fixed bulge mass for the extreme ends of the $L_{x} - M_{halo}$ plane (samples are labelled ex-XRS and ex-XRW in Fig.~\ref{fig1}) to ensure that the trends we observe with X-ray brightness are robust. In Fig.~\ref{fig8} we see that compared to our average trends, galaxies in extremely X-ray bright environments have smaller disc scale lengths, whereas in extremely X-ray faint environments galaxy disc scale lengths approach the trends for the isolated field sample. This confirms that the differences we see are not simply a result of scatter in the $L_{x} - M_{halo}$ relation, since the trends exhibit a stronger dependence on X-ray luminosity in the extremes of the dataset.

\subsection{Star-forming and quiescent discs}

At higher redshifts ($z \sim 0.44$), \cite{Kuchner2017} find that late-type galaxies in cluster environments have smaller disc sizes than in the field. At fixed bulge-to-total mass ratio ($\rm B/T = M_{bulge}/M_{\star}$) and stellar mass, they show that quiescent spirals are smaller than star-forming spirals. After comparing disc sizes of quiescent and star-forming galaxies, they find that the outer disc ``fades" in quiescent discs. They attribute these results to a combination of ram-pressure stripping and other slow gas removal processes such as starvation. However, previous work \citep[e.g.][]{Shioya2002} shows that truncated spirals only become red and quenched $\sim 1\,\mathrm{Gyr}$ after disc truncation using chemical evolution models.

\begin{figure*}
	\centering
	\begin{minipage}[b]{.45\textwidth}
	\includegraphics[width=1\textwidth]{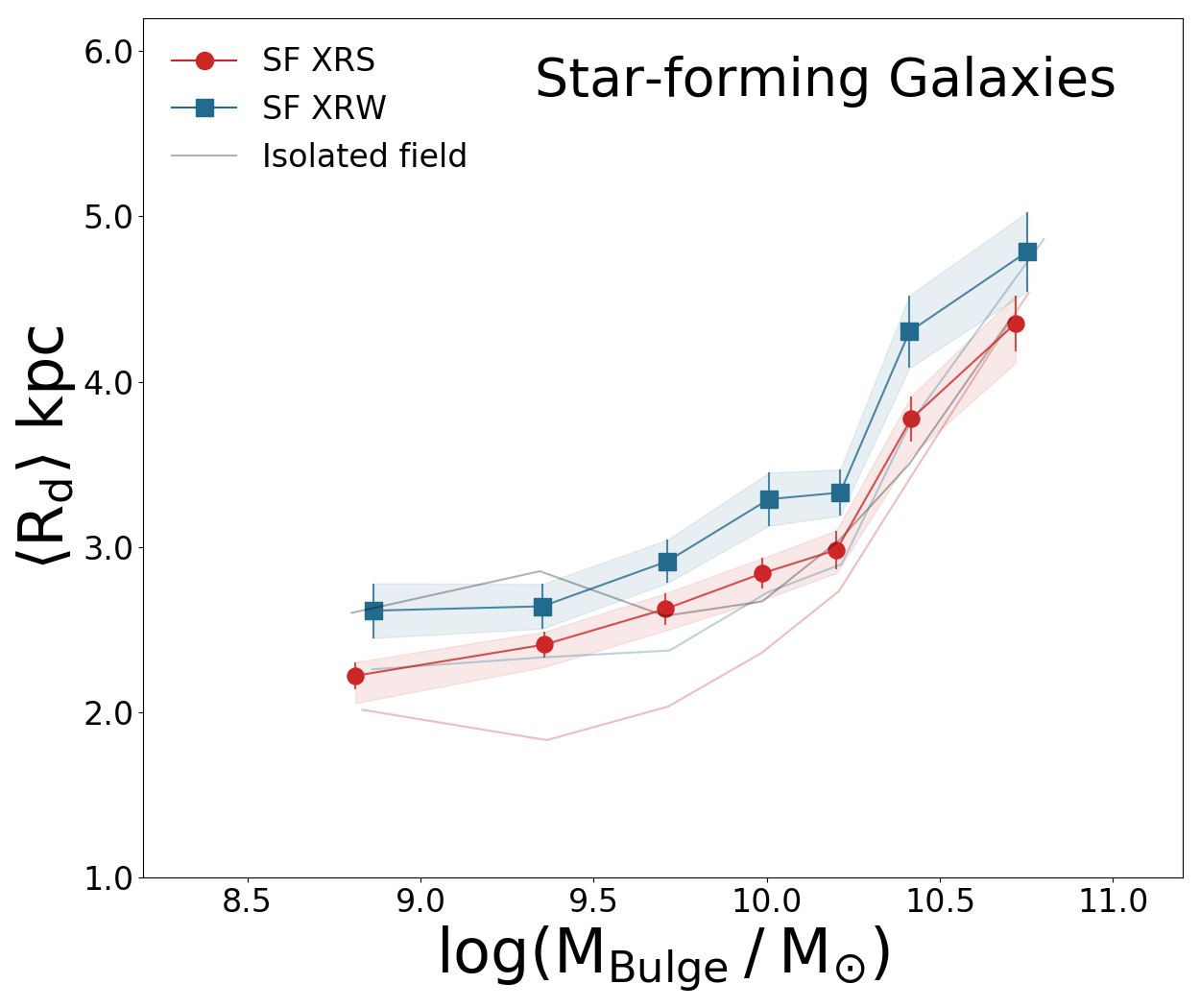}
	\caption{$V_{max}$ weighted exponential disc scale length versus bulge mass. Bold lines correspond to star-forming galaxies. Faded lines correspond to average trends for entire sample in Fig.~\ref{fig6}. Error bars and shaded regions correspond to the standard error of the mean. \newline}
	
	\label{fig9}
	\end{minipage}%
	\hspace{1cm}
	\begin{minipage}[b]{.45\textwidth}
		
		\includegraphics[width=\columnwidth]{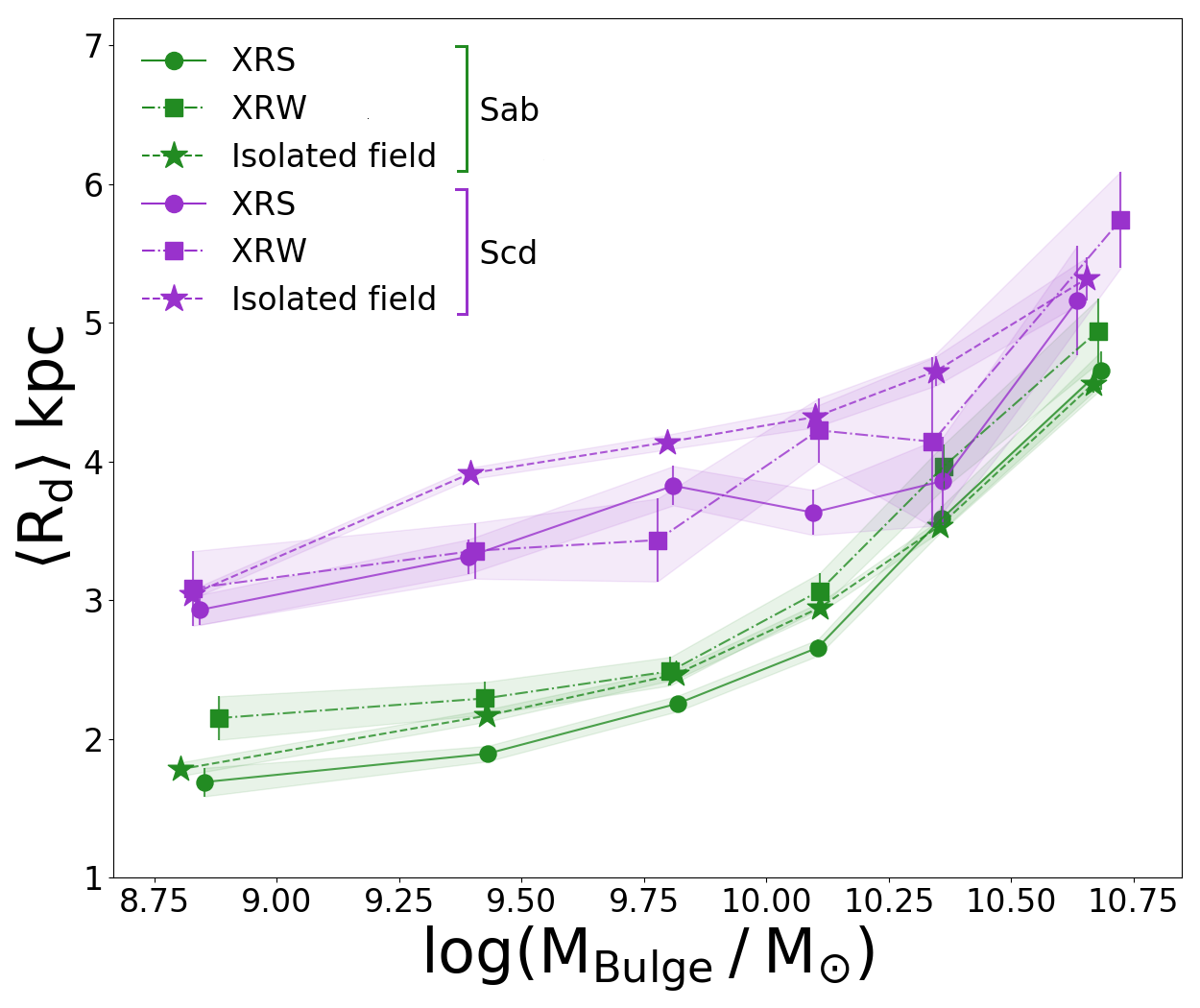}
		\caption{$V_{max}$ weighted exponential disc scale length versus bulge mass by morphological type (Sa/Sbs in green and Sc/Sds in purple). Dashed lines are isolated field galaxies, XRW are dot-dashed and XRS are solid. Error bars and shaded regions correspond to the standard error of the mean.}
		
		\label{fig10}
	\end{minipage}
\end{figure*}

To explore whether the galaxies in our samples undergo disc-fading from the outside-in, we make a simple cut in specific star-formation rate ($\rm SSFR = SFR/M_{\star}$) using star-formation rates (SFR) from \cite{Brinchmann2004} and stellar masses from \citet{Mendel2014}. The ``break" between the red quiescent and blue star-forming population of galaxies at low-redshift occurs at $\rm \log{SSFR} \sim -11$ \citep{Wetzel2012}. In Fig.~\ref{fig9} we show our results for galaxies classified as star-forming (with $\rm \log{SSFR} > -11$) overlaid on the average trends for the full sample (see Fig.~\ref{fig6}). For $\rm \log(M_{bulge}/M_{\odot}) \lesssim 10.5$, star-forming galaxies in XRS and XRW environments have significantly larger disc scale lengths than the average trends from the total sample. This suggests that as star-forming discs quench, their exponential disc scale lengths decrease from the outside-in. This is an indication that we are observing outside-in disc fading occuring in our sample.  

\subsection{Trends by morphological type}
\label{sec: 4.4}

To determine whether our trends are specific to early or late-type morphological classifications within our sample, we use the machine learning catalogue of \cite{Huertas-Company2011} that divides SDSS galaxies into four types: ellipticals (E), lenticular galaxies (S0), intermediate spirals (Sa/Sb) and late-type spirals (Sc/Sd). Since there are biases in detecting differences between E/S0 galaxies in this catalogue and we are primarily interested in discs, we remove these early-types from our analysis. In Fig.~\ref{fig10}, we see that Sc/Sds have significantly larger disc scale lengths compared to Sa/Sbs, consistent with previous work \citep[e.g.][]{Courteau2007, Vaghmare2015}. When we subdivide these morphological types by isolated field, XRS and XRW environments, we find that the disc scale lengths of spiral galaxies (Sa/Sb and Sc/Sd) at fixed bulge mass show some trends with environment though they are not as clear as in Fig.~\ref{fig6}.

\subsection{Putting it all together}

We find that galaxy disc scale lengths are sensitive to X-ray environment, a tracer of IGM density. At fixed stellar mass, Fig.~\ref{fig5} shows that galaxies in group environments have smaller disc scale lengths than isolated field galaxies. There are a number of processes that could lead to smaller disc scale lengths in groups compared to the field, but the fact that galaxies have smaller disc scale lengths in XRS groups compared to XRW groups (which are defined at fixed halo mass in \textsection \ref{sec: 2.1.2}) suggests that IGM density plays an important role. Since hydrodynamical processes are most effective in the densest IGMs, we expect environmental mechanisms such as starvation and ram-pressure stripping to be most efficient in XRS group environments. The fact that the differences in disc scale length are enhanced in the X-ray extremes of our dataset is consistent with a hydrodynamic origin. 

Morphology scaling relations at fixed stellar mass have long been shown to only weakly depend on environmental density \citep[e.g.][]{Bamford2009}, in agreement with our findings. Although, in Fig.~\ref{fig7} we do see larger differences between galaxies in XRS and XRW systems at small group-centric radii, which could be due to enhanced stripping; since these galaxies would be moving fastest, and through the densest IGMs. Detailed exploration of the possible physical mechanisms at play awaits future work. What is clear is that disc scale length is sensitive to X-ray brightness.

\section{Conclusions}
\label{sec: 5}

We study a sample galaxies from SDSS-DR7 in the \citetalias{Yang2007} group catalogue with X-ray data from \cite{Wang2014} and bulge-disc structural parameters from \citetalias{Simard2011} and \cite{Mendel2014}. We note that disc scale lengths of low mass galaxies are known to have large systematic uncertainties, however we focus on differences between our environmental samples which share the same systematics. With this is mind, we find that:

\begin{enumerate}
	\item The exponential disc scale lengths of low mass galaxies in groups (both X-ray strong and X-ray weak) are smaller than isolated field galaxies by $\sim 1\,\mathrm{kpc}$, consistent with previous work. Additionally, the disc scale lengths of low mass galaxies in XRS groups are smaller by $\sim 0.5\,\mathrm{kpc}$ compared to galaxies in XRW groups. These results are consistent when using \citet{Meert2015} disc scale length measurements.
	\item Disc scale lengths at fixed bulge mass do not show a strong dependence on halo mass. The observed differences between galaxy disc scale lengths in XRS and XRW groups are dominated by galaxies located at small group-centric radii.
	\item The differences between disc scale lengths for galaxies in XRS and XRW groups are enhanced when comparing galaxies in the most X-ray bright groups (ex-XRS) to those in the most X-ray faint groups (ex-XRW).
	\item Star-forming spiral galaxies in XRS and XRW environments have larger stellar discs than the average trends for our total group samples.

\end{enumerate}

Together these findings demonstrate the impact of X-ray environment on galaxy disc scale lengths of low mass galaxies, which should be confirmed with deeper higher image quality imaging surveys such as DECaLS \citep[see][]{Dey2019}. Future studies with resolved Integrated Field Unit (IFU) maps could be used to extract star-forming radii to compare to the disc scale lengths from traditional bulge-disc decompositions. Taken one step further, applying bulge-disc decompositions to IFU datacubes directly \citep{Johnston2017} and analyzing how galaxy structural components vary with environment could shed light on the transformation of galaxy disc morphology.

\section*{Acknowledgements}

The authors thank the anonymous referee for their helpful comments
and suggestions which have improved this paper. 
IDR and LCP thank the National Science and Engineering Research Council
of Canada for funding. MLD thanks the National Indian Brotherhood Trust for scholarship funding, Fraser Evans for his helpful comments, and Sarah Demers for her input. Our research would not be possible without access to a wealth of public datasets. We thank X. Yang et al. for making their SDSS DR7 group catalogue publicly available, L. Wang et al. for the publication of their X-ray group catalogue, L. Simard et al. for the publication of their SDSS DR7
morphology catalogue, J. T. Mendel et al. for the publication of their bulge and disc mass catalogue, A. Meert et al. for the publication of their bulge-disc decomposition catalogue, and the NYU-VAGC team for the publication of their SDSS-DR7 catalogue. This work also made use of many open-source software packages, such as: \texttt{Matplotlib} \citep{Hunter2007}, \texttt{NumPy} \citep{vanderWalt2011}, \texttt{Pandas} \citep{McKinney2010}, \texttt{SciPy} \citep{Jones2001}, and \texttt{Topcat} \citep{Taylor2005}.

Funding for the SDSS has been provided by the Alfred
P. Sloan Foundation, the Participating Institutions, the National
Science Foundation, the U.S. Department of Energy, the National
Aeronautics and Space Administration, the Japanese Monbukagakusho,
the Max Planck Society, and the Higher Education
Funding Council for England. The SDSS Web Site is
http://www.sdss.org/.

The SDSS is managed by the Astrophysical Research Consortium
for the Participating Institutions. The Participating Institutions
are the American Museum of Natural History, Astrophysical
Institute Potsdam, University of Basel, University of Cambridge,
Case Western Reserve University, University of Chicago, Drexel
University, Fermilab, the Institute for Advanced Study, the Japan
Participation Group, Johns Hopkins University, the Joint Institute
for Nuclear Astrophysics, the Kavli Institute for Particle Astrophysics
and Cosmology, the Korean Scientist Group, the Chinese
Academy of Sciences (LAMOST), Los Alamos National Laboratory,
the Max-Planck-Institute for Astronomy (MPIA), the MaxPlanck-Institute
for Astrophysics (MPA), New Mexico State University,
Ohio State University, University of Pittsburgh, University
of Portsmouth, Princeton University, the United States Naval Observatory, and the University of Washington.

%%%%%%%%%%%%%%%%%%%%%%%%%%%%%%%%%%%%%%%%%%%%%%%%%%



\bibliographystyle{mnras}
\bibliography{example}

%%%%%%%%%%%%%%%%%%%%%%%%%%%%%%%%%%%%%%%%%%%%%%%%%%
%%%%%%%%%%%%%%%%%%%%%%%%%%%%%%%%%%%%%%%%%%%%%%%%%%

%%%%%%%%%%%%%%%%% APPENDICES %%%%%%%%%%%%%%%%%%%%%

\appendix

\section{Machine readable tables}
\label{appendix: a}

We have included three machine readable tables corresponding to the galaxies contained in our XRS groups (table1.csv), XRW groups (table2.csv) and in the isolated field (table3.csv) referred to in \textsection \ref{sec: 2.1.2} and \textsection \ref{sec: 2.3.2}. A full description of each table and of the galaxy and group properties we include in the data files can be found in the readme file.

\begin{table*}
	\centering
	\caption{Sample data for galaxies from the XRS group sample referred to in \textsection \ref{sec: 2.1.2} (the same data columns are used in the XRW group sample). We refer to bulge and disc parameters from \citetalias{Simard2011} and \citetalias{Mendel2014}, group parameters from \citetalias{Yang2007}, star-formation rate (SFR) from \citet{Brinchmann2004}, and  group X-ray luminosity ($L_{x}$) from \citet{Wang2014}. A full description of the columns below can be found in the catalogue readme file for table1.csv and table2.csv.}
	\begin{tabular}{lllllllllllllll} % four columns, alignment for each
		\hline
		SDSS Object ID     & Right Ascension & Declination & ... & Disc Scale Length & ... & Group ID  & ... & Group-centric Radius \\
		objID              & RA [deg]                    & DE [deg]               & ... & $R_{d}$ [kpc]           & ... & grID   & ... & R/R200               \\
		Column 1           & Column 2              & Column 3          & ... & Column 11         & ... & Column 17 & ... & Column 21            \\
		\hline
		587722982832537906 & 222.81485             & -0.37215          & ... & 2.96              & ... & 1349      & ... & 0.350096             \\
		587722982832668754 & 223.04597             & -0.2561           & ... & 5.24              & ... & 1349      & ... & 0.557528             \\
		587722982832668797 & 223.06693             & -0.29362          & ... & 3.31              & ... & 1349      & ... & 0.549946             \\
		587722982832668823 & 223.0809              & -0.26921          & ... & 1.0               & ... & 1349      & ... & 0.629379             \\
		587722982832669003 & 223.08322             & -0.22582          & ... & 2.89              & ... & 1349      & ... & 0.719030            
	
	\end{tabular}
\label{tab: tab1}
\end{table*}

\begin{table*}
	\centering
	\caption{Sample of galaxies from the isolated field catalogue referred to in \textsection \ref{sec: 2.3.1}. We refer to bulge and disc parameters from \citetalias{Simard2011} and \citetalias{Mendel2014}, star-formation rate (SFR) from \citet{Brinchmann2004}, and we reference the \citetalias{Yang2007} group ID for single group galaxies in this sample. A full description of the columns below can be found in the catalogue readme file for table3.csv.}
	\begin{tabular}{lllllllllllllll} % four columns, alignment for each
		\hline
		SDSS Object ID     & Right Ascension & Declination         & ... & Disc Scale Length & ... & Group ID  & Star-formation Rate \\
		objID              & RA [deg]    & DE [deg]        & ... & $R_{d}$ [kpc] & ... & grID      & SFR [$M_{\odot}yr^{-1}$]          \\
		Column 1           & Column 2        & Column 3            & ... & Column 11         & ... & Column 15 & Column 16           \\
		\hline
		587722953840787640 & 236.30992       & 0.77218             & ... & 2.87              & ... & 560811    & 1.0815247           \\
		587722981750931635 & 204.80522       & -1.12092 & ... & 1.88              & ... & 38796     & -0.43257746         \\
		587722982281445440 & 190.30172       & -0.75682 & ... & 1.03              & ... & 41258     & -0.97517890 \\
		587722982822838494 & 200.66116       & -0.28191            & ... & 2.64              & ... & 45213     & -0.99877715         \\
		587722982823624834 & 202.35513       & -0.39903            & ... & 2.12              & ... & 45276     & -1.7329755       
	\end{tabular}
\label{tab: tab2}
\end{table*}

\section{Comparison to Meert et al. 2015}
\label{appendix: b}

Here we compare our measurements based on structural parameters found in \citetalias{Simard2011} to those of \citet{Meert2015}. \citet{Meert2015} provide independent photometric decomposition catalogues for four models: a pure S\'ersic model (denoted Ser), a de Vaucouleurs profile (deV), a  S\'ersic bulge + exponential disc model (Ser-Exp), and a de Vaucouleurs + exponential disc model (deV-Exp). They perform these decompositions in the \textit{g}, \textit{r}, and \textit{i} bandpasses. For the best comparison to our work with \citetalias{Simard2011} table 1 (a de Vaucouleurs bulge and exponential disc), in this appendix we compare to the \citet{Meert2015} dev-Exp \textit{r}-band model results.

\begin{figure}
	\includegraphics[width=\columnwidth]{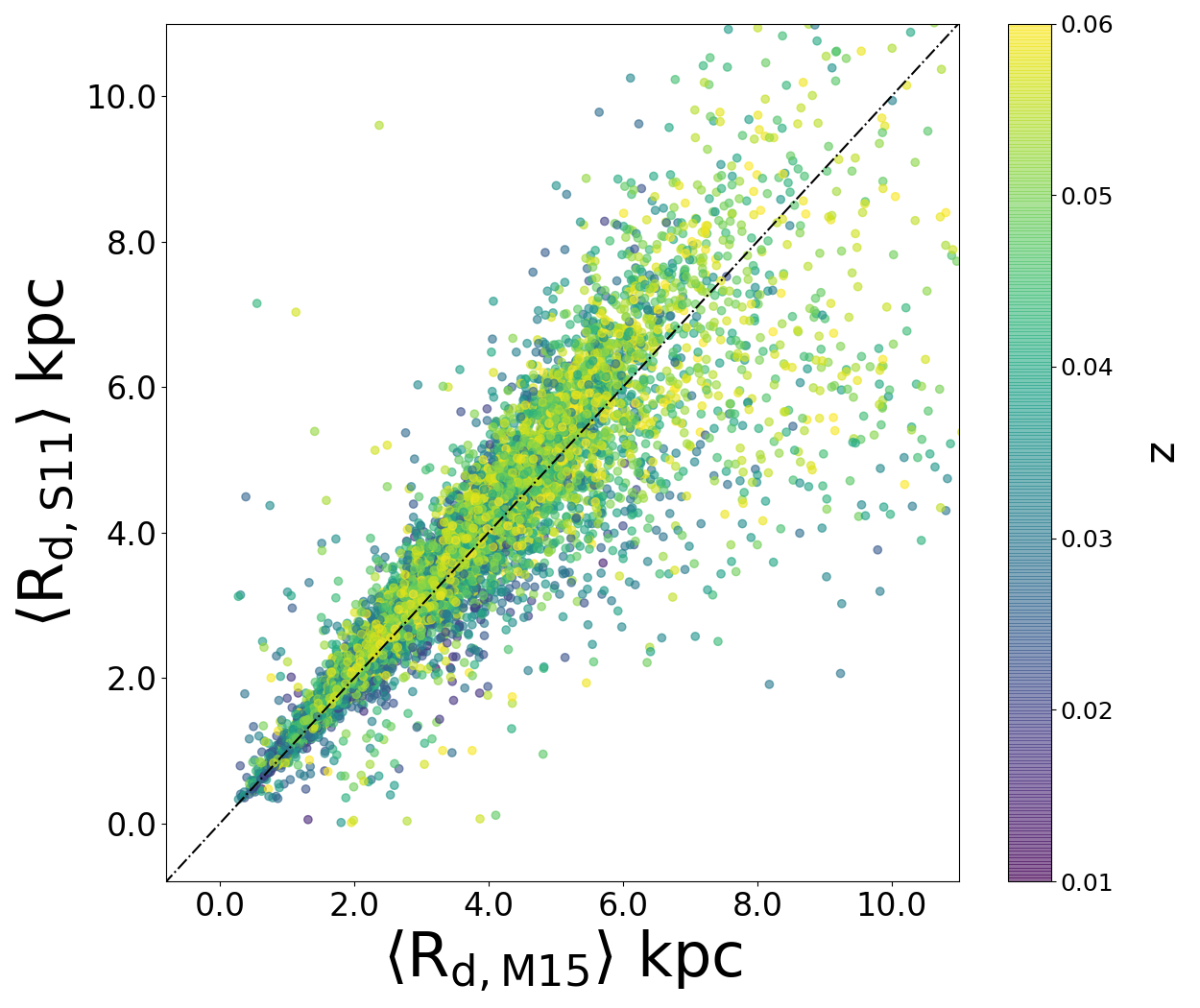}
	\caption{$R_{d, S11}$ versus $R_{d, M15}$ colour coded by redshift for all environment samples combined. The dashed line corresponds to a 1:1 linear relationship. The data in this figure have been trimmed at $R_{d} = 11.0\, \mathrm{kpc}$ on both axes. A small number of objects had larger measured disc scale lengths in both catalogues.}
	
	\label{fig11}
\end{figure}

Matching the deV-Exp \textit{r}-band \citet{Meert2015} galaxies to our XRS, XRW and isolated field samples yields sample sizes of 3,268 XRS galaxies, 1,144 XRW galaxies and 10,307 isolated field galaxies, somewhat smaller than the \citetalias{Simard2011} samples described in \textsection \ref{sec: 2.3}. In Fig.~\ref{fig11} the exponential disc scale length from \citetalias{Simard2011} is compared to the exponential disc scale length derived from \citet{Meert2015}, given by the disc half-light radius $R_{disc, hl} / 1.7$.

In Fig.~\ref{fig11}, we show the exponential disc scale length from \citetalias{Simard2011} versus the exponential disc scale length from \citet{Meert2015}, for our entire sample of XRS, XRW and isolated galaxies merged into one dataset. We find that there is good correlation between disc scale length measurements from the two catalogues. We also note that there is no strong redshift dependence.

\begin{figure}
	\includegraphics[width=\columnwidth]{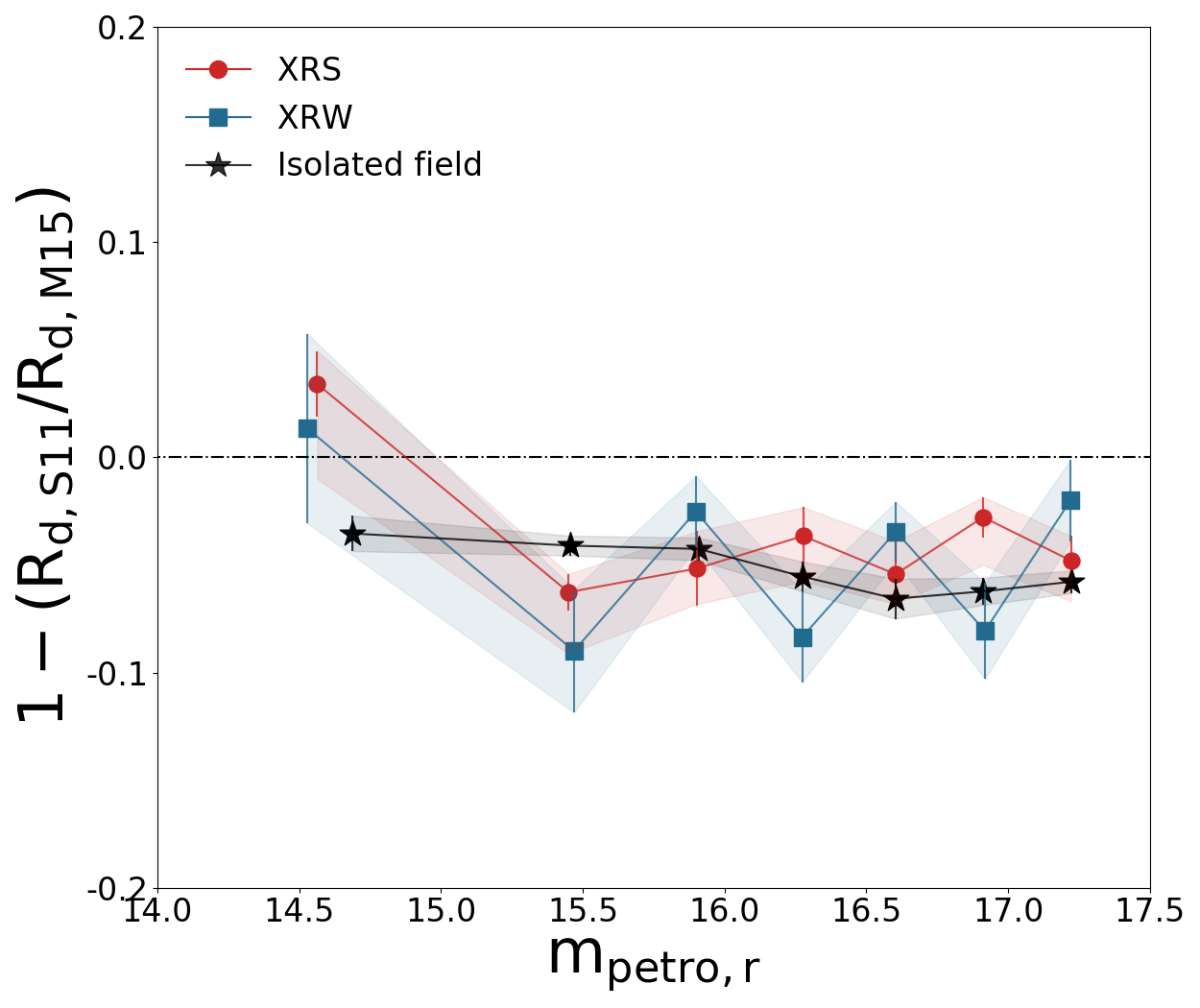}
	\caption{Ratio of exponential disc scale lengths from \citetalias{Simard2011} and \citet{Meert2015} in our matched sample versus \textit{r}-band apparent magnitude from \citet{Meert2015}. This figure agrees with the figure provided in Appendix C3, Figure C4, row 4 of \citet{Meert2015}.}
	
	\label{fig13}
\end{figure}

In Appendix C3 from \citet{Meert2015}, the authors compare the results of their deV-Exp bulge-disc decompositions to \citetalias{Simard2011} and \citet{LacknerGunn2012}. In Fig.~\ref{fig13} we repeat this analysis for our matched sample of galaxies from \citetalias{Simard2011} and \citet{Meert2015} (with $z$, $M_{\star}$ and $P_{pS}$ cuts applied). The disc scale lengths measured in \citetalias{Simard2011} are similar but systematically larger than those of \citet{Meert2015} at faint magnitudes. We find good agreement between Fig.~\ref{fig13} for our \citetalias{Simard2011}-\citet{Meert2015} matched sample and the results shown in \citet{Meert2015} in Appendix C3, Figure C4, row 4 where the authors also compare their disc scale length results to \citetalias{Simard2011}.

%%%%%%%%%%%%%%%%%%%%%%%%%%%%%%%%%%%%%%%%%%%%%%%%%%
%%%%%%%%%%%%%%%%%%%%%%%%%%%%%%%%%%%%%%%%%%%%%%%%%%

% Don't change these lines
\bsp	% typesetting comment
\label{lastpage}
\end{document}